\documentclass
[aps,pra,reprint,showpacks,twocolumn,superscriptaddress]{revtex4-1}%
\usepackage[colorlinks,linkcolor=blue,citecolor=blue,hyperindex,bookmarks=false,pdfstartview=FitH]%
{hyperref}
\usepackage{amsfonts}
\usepackage{mathrsfs}
\usepackage{amsmath}
\usepackage{amssymb}
\usepackage{graphicx}
\usepackage{dcolumn}
\usepackage{color}
\usepackage{bm}
\usepackage{soul}
\usepackage{subfigure}
\usepackage[colorlinks,linkcolor=blue,citecolor=blue,hyperindex,bookmarks=false,pdfstartview=FitH]%
{hyperref}%
\providecommand{\U}[1]{\protect\rule{.1in}{.1in}}
\providecommand{\U}[1]{\protect\rule{.1in}{.1in}}

\providecommand{\U}[1]{\protect\rule{.1in}{.1in}}
\providecommand{\U}[ 1]{\protect\rule{.1in}{.1in}}
\providecommand{\U}[1]{\protect\rule{.1in}{.1in}}
\providecommand{\U}[1]{\protect\rule{.1in}{.1in}}

\providecommand{\U}[1]{\protect\rule{.1in}{.1in}}

\begin{document}
\title{Unidirectional Photonic Reflector Using a Defective Atomic Lattice}
\author{Tianming Li}
\affiliation{School of Physics and Electronic Engineering, Hainan Normal University, Haikou
571158, P. R. China}
\affiliation{School of Physics and Telecommunication Engineering, South China
Normal University, Guangzhou 510006, P. R. China}
\author{Hong Yang }
\email{yang_hongbj@126.com}
\affiliation{School of Physics and Electronic Engineering, Hainan Normal University, Haikou
571158, P. R. China}
\author{Maohua Wang}
\affiliation{Center for Quantum Sciences and School of Physics, Northeast Normal
University, Changchun 130024, P. R. China}
\author{Chengping Yin}
\affiliation{School of Physics and Telecommunication Engineering, South China
Normal University, Guangzhou 510006, P. R. China}
\author{Tinggui Zhang }
\email{tinggui333@163.com}
\affiliation{School of Mathematics and Statistics, Hainan Normal University, Haikou
571158, P. R. China}
\author{Yan Zhang}
\email{zhangy345@nenu.edu.cn}
\affiliation{Center for Quantum Sciences and School of Physics, Northeast Normal
University, Changchun 130024, P. R. China}
\date{\today}

\begin{abstract}
Based on the broken spatial symmetry, we propose a novel scheme for
engineering a unidirectional photonic reflector using a one-dimensional atomic
lattice with defective cells that have been specifically designed to be
vacant. By trapping three-level atoms and driving them into the regime of
electromagnetically induced transparency, and through the skillful design of
the number and position of vacant cells in the lattice, numerical simulations
demonstrate that a broad and high unidirectional reflection region can be
realized within EIT window. This proposed unidirectional reflector scheme
provides a new platform for achieving optical nonreciprocity and has potential
applications for designing optical circuits and devices of nonreciprocity at
extremely low energy levels.

\end{abstract}
\maketitle

\affiliation{School of Physics and Electronic Engineering, Hainan Normal University, Haikou
571158, P. R. China}
\affiliation{School of Physics and Telecommunication Engineering, South China
Normal University, Guangzhou 510006, P. R. China}

\affiliation{School of Physics and Electronic Engineering, Hainan Normal University, Haikou
571158, P. R. China}

\affiliation{Center for Quantum Sciences and School of Physics, Northeast Normal
University, Changchun 130024, P. R. China}

\affiliation{School of Physics and Telecommunication Engineering, South China
Normal University, Guangzhou 510006, P. R. China}

\affiliation{School of Mathematics and Statistics, Hainan Normal University, Haikou
571158, P. R. China}

\affiliation{Center for Quantum Sciences and School of Physics, Northeast Normal
University, Changchun 130024, P. R. China}

\affiliation{School of Physics and Electronic Engineering, Hainan Normal University, Haikou
571158, P. R. China}
\affiliation{School of Physics and Telecommunication Engineering, South China
Normal University, Guangzhou 510006, P. R. China}

\affiliation{School of Physics and Electronic Engineering, Hainan Normal University, Haikou
571158, P. R. China}

\affiliation{Center for Quantum Sciences and School of Physics, Northeast Normal
University, Changchun 130024, P. R. China}

\affiliation{School of Physics and Telecommunication Engineering, South China
Normal University, Guangzhou 510006, P. R. China}

\affiliation{School of Mathematics and Statistics, Hainan Normal University, Haikou
571158, P. R. China}

\affiliation{Center for Quantum Sciences and School of Physics, Northeast Normal
University, Changchun 130024, P. R. China}

\affiliation{School of Physics and Electronic Engineering, Hainan Normal University, Haikou
571158, P. R. China}
\affiliation{School of Physics and Telecommunication Engineering, South China
Normal University, Guangzhou 510006, P. R. China}

\affiliation{School of Physics and Electronic Engineering, Hainan Normal University, Haikou
571158, P. R. China}

\affiliation{Center for Quantum Sciences and School of Physics, Northeast Normal
University, Changchun 130024, P. R. China}

\affiliation{School of Physics and Telecommunication Engineering, South China
Normal University, Guangzhou 510006, P. R. China}

\affiliation{School of Mathematics and Statistics, Hainan Normal University, Haikou
571158, P. R. China}

\affiliation{Center for Quantum Sciences and School of Physics, Northeast Normal
University, Changchun 130024, P. R. China}

\affiliation{School of Physics and Electronic Engineering, Hainan Normal University, Haikou
571158, P. R. China}
\affiliation{Guangdong Provincial Key Laboratory of Quantum Engineering and Quantum
Materials, School of Physics and Telecommunication Engineering, South China
Normal University, Guangzhou 510006, P. R. China}

\affiliation{School of Physics and Electronic Engineering, Hainan Normal University, Haikou
571158, P. R. China}

\affiliation{Center for Quantum Sciences and School of Physics, Northeast Normal
University, Changchun 130024, P. R. China}

\affiliation{Guangdong Provincial Key Laboratory of Quantum Engineering and Quantum
Materials, School of Physics and Telecommunication Engineering, South China
Normal University, Guangzhou 510006, P. R. China}

\affiliation{School of Mathematics and Statistics, Hainan Normal University, Haikou
571158, P. R. China}

\affiliation{Center for Quantum Sciences and School of Physics, Northeast Normal
University, Changchun 130024, P. R. China}

In order to achieve the practical applications of nonreciprocal photonic
devices and circuits without the use of magneto-optical materials, which can
hinder miniaturization and integration, such as chip isolators, circulators,
and all-optical diodes \cite{device-1, device-2, device-3, device-4, device-5,
device-6}, the research on controllable unidirectional light propagation to
induce a symmetry-breaking effect is of great significance within the science
and engineering disciplines \cite{SE-1, SE-2, SE-4, SE-5}. Various strategies
for achieving optical nonreciprocity have been explored, such as breaking the
time-reversal symmetry \cite{timeb} in atom-cavity systems \cite{YPF},
optomechanical systems \cite{opto-1, opto-2, opto-3, opto-4}, magnomechanical
systems \cite{magn-1, magn-2}, giant-atom systems \cite{timeb}, quantum
squeezing resonator modes in chip systems \cite{xia}, and considering Doppler
shifts in hot atomic systems \cite{hot-1, hot-2, hot-3}.

In recent years, significant progress has been made in the study of
unidirectional reflection (Here, the unidirectional reflection, i.e., the
perfectly asymmetric reflection, indicates that the reflection of the wave
incident from one side is completely suppressed when the wave incident from
the other side is typically reflected.) of nonlinear materials in various
physical systems. For instance, by trapping atoms in space with a moving
optical lattice, a probe pulse propagating along the lattice from two sides
could experience nonreciprocal reflection \cite{moving-1, moving-2, moving-3};
however, the implementation of moving lattices requires relatively high
technological requirements. For a static optical lattice, by skillfully
coupling it with standing-wave fields for a dynamic frequency shift as a
cosine/sine function of the lattice position, the initially uniform probe
susceptibility can be modulated to satisfy optical parity-time (PT) symmetry
(antisymmetric) \cite{PT-1, PT-2, PT-3}, leading to unidirectional
reflectionless. In addition, unidirectional reflection can be achieved in
continuous media when the refractive index obeys the spatial Kramers-Kronig
(KK) relations \cite{KK-1, KK-2, KK-3}. However, the main difficulty in
implementing these schemes lies in the need for precise modulation, including
achieving a balance of gain and loss within a single period, precisely
arranging the spatial coupling fields, skillfully constructing spatial-KK
media, and improving the low reflectivity of unidirectional reflection.

As one of the approaches to achieving a photonic bandgap (PBG) \cite{PBG-01,
PBG-02}, the use of an atomic lattice has been employed to attain a large
Bragg reflection \cite{PBG-1, PBG-2, PBG-3}. In particular, by driving
periodically trapped atoms into the regime of electromagnetically induced
transparency (EIT), we can significantly enhance the reflectivity and achieve
additional PBGs and greater tunability \cite{PBG-4, PBG-5}. Furthermore, such
reciprocal reflection is robust against disorder in the atomic lattice while
maintaining periodicity \cite{Yang2020PRA}, but can be disrupted in a
defective atomic lattice as we found \cite{Li2021OE}. Maybe, distinct and
intriguing reflection phenomena will be observed, in a purposely designed
defective lattice.

\begin{figure}[t]
\includegraphics[width=0.5\textwidth]{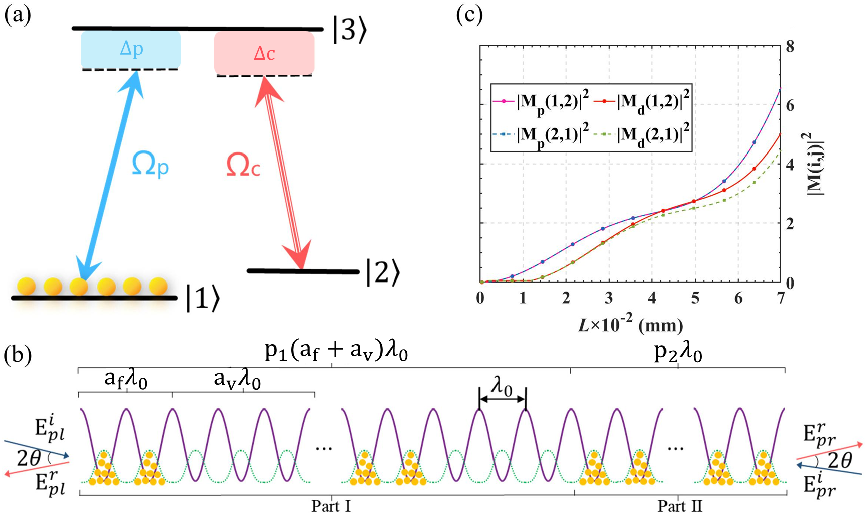}\caption{(color online) (a)
Energy level diagram of a three-level $\Lambda$-type atomic system interacting
with a weak probe field $\Omega_{p}$ and a strong coupling field $\Omega_{c}$.
(b) 1D defective atomic lattice with the width $\lambda_{0}$ for each period,
which comprises Part I and Part II. The probe filed $\mathbf{E}_{p}$ is
incident from either the left side (denoted by $\mathbf{E}_{pl}^{i}$) or the
right side (denoted by $\mathbf{E}_{pr}^{i}$) with a small incident angle
$\theta$ relative to $x$-axis. The relevant reflected fields are denoted by
$\mathbf{E}_{pl}^{r}$ and $\mathbf{E}_{pr}^{r}$. (c) The Square of moduli of
matrix elements $M_{p,d}(1,2)$, $M_{p,d}(2,1)$ as functions of length $L$, for
a perfect lattice with $180$ fulfilled cells and a defective lattice with
$a_{f}=5$, $a_{v}=25$, $p_{1}=1$, $p_{2}=150$, respectively. The parameters
are $\Gamma_{31}=\Gamma_{21} =6.0$ MHz, $\Omega_{c}=36.0$ MHz, $\Delta
_{c}=15.0$ MHz, $\lambda_{Lat}=781.3$ nm, $\lambda_{31}=780.24$ nm,
$\Delta\lambda_{Lat}=0.9$ nm, $\eta=5.0$, $N_{0}=7.0\times10^{11}$ cm$^{-3}$,
and $d_{23}=1.0357\times10^{-29}$ Cm.}%
\label{fig1}%
\end{figure}

In this study, we propose a novel scheme to achieve unidirectional reflection
using a one-dimensional (1D) defective atomic lattice, comprising both filled
lattice cells (trapping atoms) and vacant cells (not trapping atoms). Firstly,
we drive the atoms into a three-level $\Lambda$-type EIT system [see
Fig.\ref{fig1}(a)]. Subsequently, we engineer the lattice into two parts of
periodic structures: Part I with $p_{1}$ periods, each consisting of $a_{f}$
filled cells and $a_{v}$ vacant cells, and Part II with $p_{2}$ periods of
solely filled cells [see Fig.\ref{fig1}(b)]. The total length of the lattice
is $L=S\lambda_{0}$, where $S=(a_{f}+a_{v})p_{1}+p_{2}$. Our proposal utilizes
vacant cells to break spatial symmetry, inducing unidirectional reflection
with high reflectivity in the EIT window.

We consider cold $^{87}$Rb atoms that are trapped and driven by two coherent
fields, as depicted in Fig.~\ref{fig1}(a). The weak probe field with frequency
$\omega_{p}$ (amplitude $\mathbf{E}_{{p}}$), and the strong coupling field
with frequency $\omega_{{c}}$ (amplitude $\mathbf{E}_{c}$). The fields
$\omega_{p}$ and $\omega_{c}$ interact via the dipole-allowed transitions
$\left\vert 1\right\rangle \leftrightarrow\left\vert 3\right\rangle $ and
$\left\vert 2\right\rangle \leftrightarrow\left\vert 3\right\rangle $,
respectively. The Rabi frequencies (detunings) for the probe and coupling
fields are denoted by $\Omega_{p}=\mathbf{E}_{{p}}\cdot\mathbf{d}_{13}/2\hbar$
($\Delta_{p}=\omega_{31}-\omega_{p}$) and $\Omega_{c}=\mathbf{E}_{{c}}%
\cdot\mathbf{d}_{23}/2\hbar$ ($\Delta_{c}=\omega_{32}-\omega_{c}$), where
$\mathbf{d}_{ij}$ and $\omega_{ij}$ are the dipole moments and resonant
frequencies for the relevant transitions, respectively. 
The levels $\left\vert
1\right\rangle $, $\left\vert 2\right\rangle $, and $\left\vert 3\right\rangle
$, respectively, could correspond to $\left\vert 5S_{1/2},F=1,m_{F}%
=1\right\rangle $, $\left\vert 5S_{1/2},F=2,m_{F}=2\right\rangle $, and
$\left\vert 5P_{3/2},F=2,m_{F}=2\right\rangle $ in D2 line of $^{87}$Rb atoms.
Thus, the probe field is $\sigma^{+}$ polarized, while the
coupling field is $\pi$ polarized.

Within the rotating wave and electric dipole approximations, the interaction
Hamiltonian of the system can be expressed as
\begin{align}
H_{I}  &  =\hbar\lbrack(\Delta_{p}-\Delta_{c})\left\vert 2\right\rangle
\left\langle 2\right\vert +\Delta_{p}\left\vert 3\right\rangle \left\langle
3\right\vert ]\nonumber\\
&  -\hbar\lbrack\Omega_{p}\left\vert 3\right\rangle \left\langle 1\right\vert
+\Omega_{c}\left\vert 3\right\rangle \left\langle 2\right\vert +h.c]\text{.}
\label{Eq1}%
\end{align}
The steady-state probe susceptibility $\chi(x)$ of each filled cell can be
obtained by solving the density matrix equation of $\Lambda$-type atoms, given
by
\begin{equation}
\chi(x)=\frac{N(x)\mathbf{d}_{13}^{2}}{\hbar\varepsilon_{0}}\frac{1}{\zeta}
\text{,} \label{Eq2}%
\end{equation}
where $\zeta=\frac{\left\vert \Omega_{c}\right\vert ^{2}}{\Delta_{p}%
-\Delta_{c}-i\gamma_{21}}-i\gamma_{31}$. Here, $\gamma_{ij}=(\Gamma_{i}%
+\Gamma_{j})/2$ are the coherence dephasing rates with the relevant population
decay rates $\Gamma_{i}$ and $\Gamma_{j}$. The spatial atomic density $N(x)$,
which can be considered as a Gaussian distribution in each filled cell, is
given by
\begin{equation}
N(x)=\frac{N_{0}\lambda_{0}}{\sigma_{x}\sqrt{2\pi}}\cdot e^{\left[
-(x-x_{0})^{2}/2\sigma_{x}^{2}\right]  }\text{,} \label{Eq3}%
\end{equation}
where $N_{0}$ is the average atomic density, $x_{0}$ is the trap center,
$\sigma_{x}=\lambda_{Lat}/(2\pi\sqrt{\eta})$ is the half-width with
$\eta=2U_{0}/(\kappa_{B}T)$ related to the capture depth $U_{0}$ and
temperature $T$. $\lambda_{0}=\lambda_{Lat}/2$ is the width of each cell, with
$\lambda_{Lat}$ being the wavelength of a red-detuned retroreflected laser
beam forming the optical lattice. For the Bragg condition, the incident angle
is given by $\theta=\arccos(\lambda_{p}/\lambda_{Lat0})$, where $\lambda
_{Lat0}=\lambda_{Lat}-\Delta\lambda_{Lat}$ with the geometric Bragg shift
$\Delta\lambda_{Lat}$. It is worth emphasizing that the condition for trapping
atoms is $\lambda_{Lat}>\lambda_{31}$ with $\lambda_{31}$ being the wavelength
of the transition $\left\vert 3\right\rangle $ $\leftrightarrow$ $\left\vert
1\right\rangle $.

The reflection and transmission properties of the defective lattice can be effectively characterized by a $2\times2$ unimodular transfer matrix \cite{matrix}. 
For filled lattice cells, we divide each cell into sufficient thin layers so that each layer can be regarded as homogeneous \cite{matrix2, matrix1}. 
The transfer matrix $m(x_{j})$ of a single layer can be expressed with the transmission and reflection coefficient as $t(x_{j})=1/m(2,2)$, $r_{_{l}}(x_{j})=m(1,2)/m(2,2)$ [$r_{_{r}}(x_{j})=m(2,1)/m(2,2)$]\ on the left (right) side of this layer. 
These coefficients are dependent on the Fresnel coefficients, layer thickness, and the layer's reflective index $n_{f}(x)=\sqrt{1+\chi(x)}$ \cite{book, PRA2015}. 
For the homogeneous and thin layer, $r_{_{l}}(x_{j})=r_{_{r}}(x_{j})\equiv r(x_{j})$.
Thus, $m(x_{j})$ can be written as
\begin{equation}
m(x_{j})=\frac{_{1}}{t(x_{j})}\left[
\begin{array}
[c]{cc}%
(t(x_{j})^{2}-r(x_{j})^{2}) & r(x_{j})\\
-r(x_{j}) & 1
\end{array}
\right]  \text{.}\label{Eq4}%
\end{equation}
Here, $t(x_{j})$ and $r(x_{j})$ are the transmission and reflection
coefficients of $j$th layer in each lattice cell with $j$ $\in$ $[1,100]$. The
transfer matrix of one filled cell can be expressed as
\begin{equation}
M_{f}=\Pi_{j=1}^{100}m(x_{j})\text{.}\label{Eq5}%
\end{equation}
The refractive index $n_{v}\equiv1$ for vacant lattice cells. Thus, the
transfer matrix of a vacant cell can be written as
\begin{equation}
M_{v}=\frac{_{1}}{t(x)}\left[
\begin{array}
[c]{cc}%
t(x)^{2} & 0\\
0 & 1
\end{array}
\right]  =\left[
\begin{array}
[c]{cc}%
e^{ik\lambda_{0}} & 0\\
0 & e^{-ik\lambda_{0}}%
\end{array}
\right]  \text{.}\label{Eq6}%
\end{equation}
Then, the total transfer matrix is expressed as
\begin{equation}
M_{d}=[(M_{f})^{a_{f}}\ast(M_{v})^{a_{v}}]^{p_{1}}\ast(M_{f})^{p_{2}%
}.\label{Eq7}%
\end{equation}
For a perfect atomic lattice with the same length $L$, where each cell traps
atoms with an identical distribution as Eq.~\ref{Eq3}, the total matrix can be
written as
\begin{equation}
M_{p}=\Pi_{k=1}^{S}M_{f}.\label{Eq8}%
\end{equation}
It is evident that this lattice exhibits the same reflectivities on both sides
\cite{PBG-5}.

In further, we can calculate the reflection coefficient $r_{_{l}}$ ($r_{_{r}}%
$) and transmission coefficient $t$ on the left (right) of the lattice, which
are given by
\begin{equation}
r_{_{l}}=\frac{M_{i}(1,2)}{M_{i}(2,2)}\text{, }r_{_{r}}=\frac{M_{i}%
(2,1)}{M_{i}(2,2)}\text{, }t=\frac{1}{M_{i}(2,2)}\text{,} \label{Eq9}%
\end{equation}
where $i=p,d$. Then, the relevant transmissivity $T=\left\vert t\right\vert
^{2}$, reflectivities $R_{l,r}=\left\vert r_{l,r}\right\vert ^{2}$ and
absorptivities $A_{l,r}=1-T-R_{l,r}$.

Using Eqs.~(\ref{Eq7}) and (\ref{Eq8}), we can predict and compare the
reciprocity of the probe reflection between the defective and perfect
lattices, with $\left|  M_{d,p}(1,2)\right|  ^{2}$ and $\left|  M_{d,p}%
(2,1)\right|  ^{2}$ as functions of length $L$ as shown in Fig.~\ref{fig1}(c).
While $\left|  M_{p}(1,2)\right|  \equiv\left|  M_{p}(2,1)\right|  $, $\left|
M_{d,p}(1,2)\right|  $ and $\left|  M_{d,p}(2,1)\right|  $ gradually separate
as $L$ increases. This implies that the reflection reciprocity on both sides
could be broken by eliminating atoms in some cells to induce the breaking of
spatial symmetry for the lattice, which can be observed by assessing the
reflectivities $R_{l,r}$.

\begin{figure}[ptb]
\includegraphics[width=0.5\textwidth]{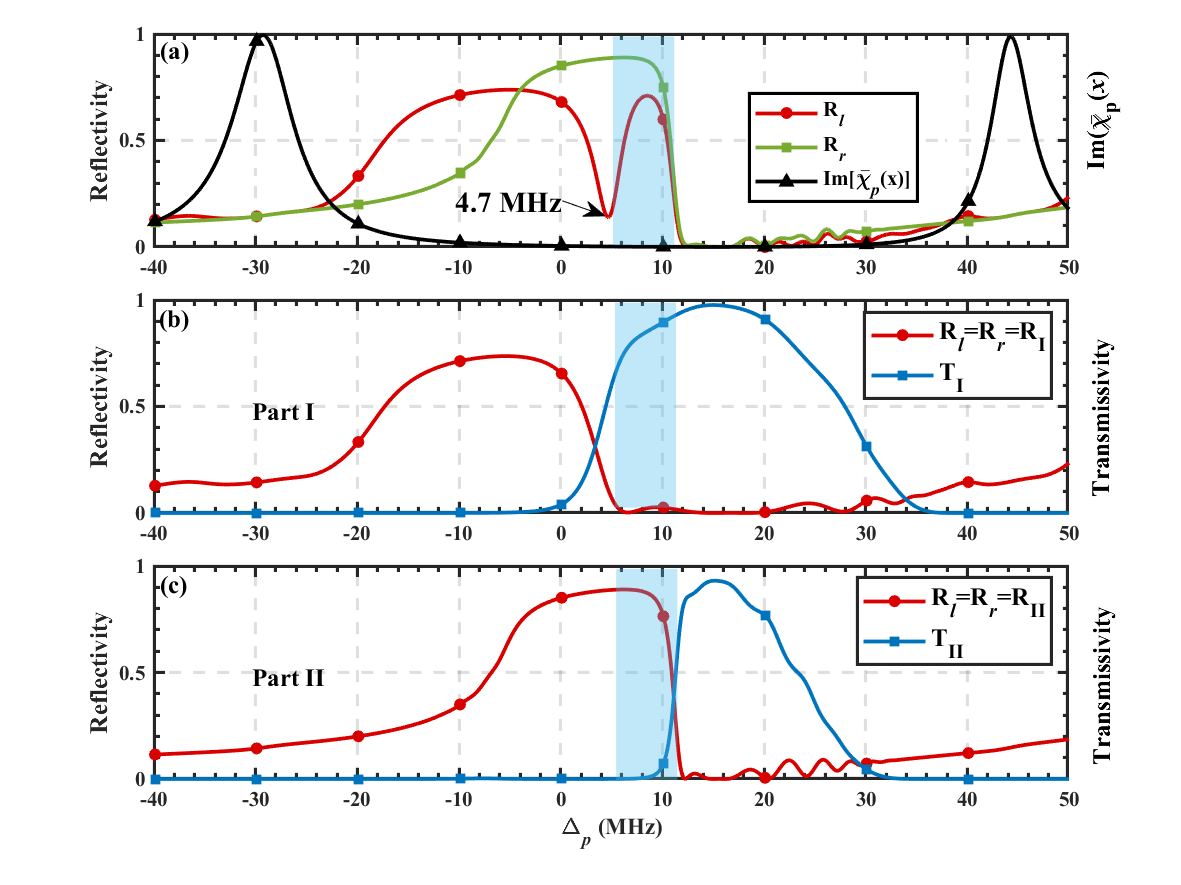}\caption{(Color online) (a)
Reflectivities $R_{l,r}$ and imaginary part of average probe susceptibility
$\overline{\chi(x)}$ versus probe detuning $\Delta_{p}$ in the defective
atomic lattice. (b) Reflectivity $R_{\text{I}}$ and transmissivity
$T_{\text{I}}$ for an individual lattice of Part I; (c) Reflectivity
$R_{\text{II}}$ and transmissivity $T_{\text{II}}$ for an individual lattice
of Part II versus detuning $\Delta_{p}$. Here $a_{f}=50$, $a_{v}=100$,
$p_{1}=10$ and $p_{2}=1500$, and other parameters are the same as
Fig.~\ref{fig1}.}%
\label{fig2}%
\end{figure}

Fig.~\ref{fig2}(a) illustrates the deviation of the left-side (LS) and
right-side (RS) reflectivities $R_{l,r}$, resulting in the reflection
nonreciprocity in the defective atomic lattice with the lengths of Part I and
Part II are equal to $0.585$ mm and then $L=1.17$ mm. The high reflectivities
of both sides are located within the EIT window $\Delta_{p}\in$($-20$ MHz,
$40$ MHz), where atomic absorption is suppressed, indicated by the imaginary
part of the average susceptibility $\overline{\chi(x)}$ near zero. At
$\Delta_{p}=4.70$ MHz, the reflectivity deviation reaches a maximum
corresponding to $R_{l}=13.90\%$ and $R_{r}=88.66\%$. While the RS reflection
band is continuous, the LS reflection is divided by the valley into a wide
platform and a narrow peak in the shaded region.

\begin{figure}[ptb]
\includegraphics[width=0.5\textwidth]{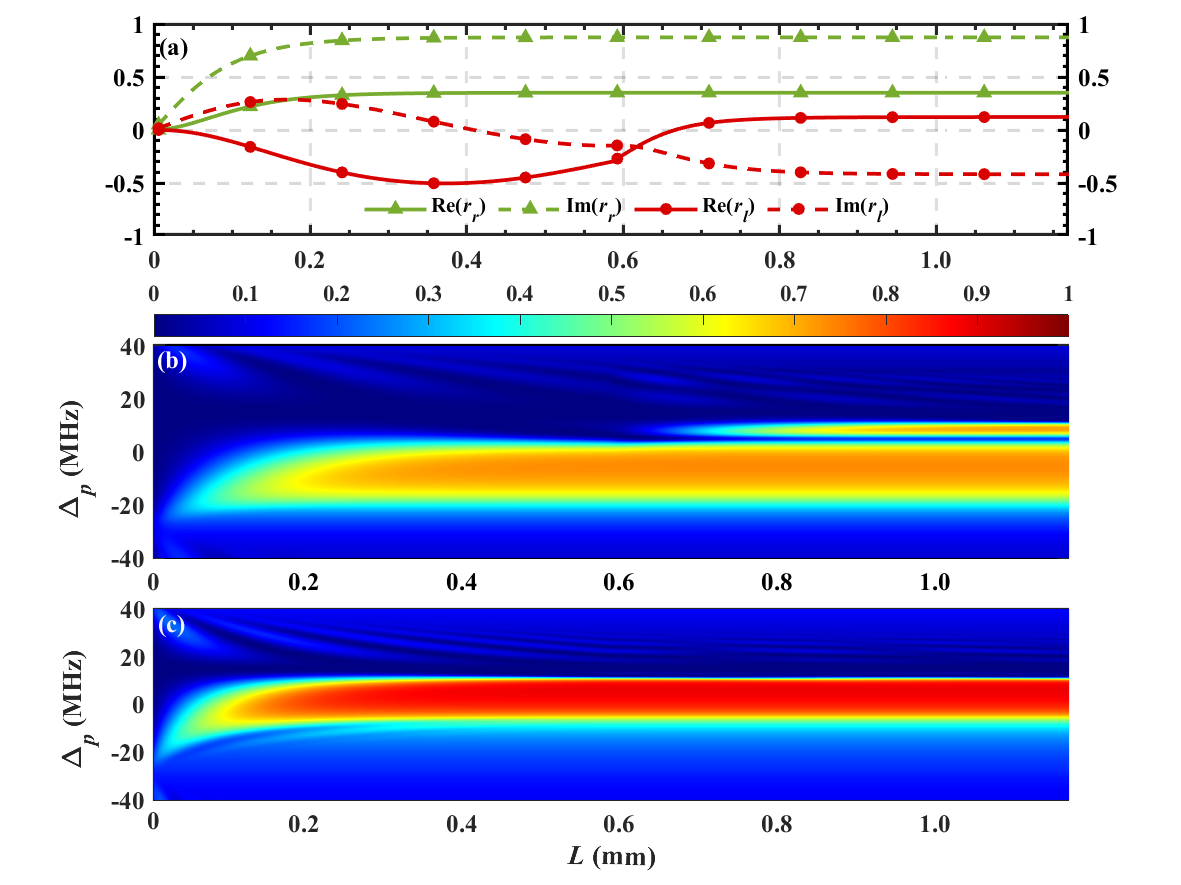}\caption{(Color online) (a)
Reflection coefficients $r_{l,r}$ versus length $L$ with $\Delta_{p}=4.7$ MHz.
(b) Reflectivity $R_{l}$ and (c) reflectivity $R_{r}$ as functions of probe
detuning $\Delta_{p}$ and length $L$. The parameters are the same as
Fig.~\ref{fig2}.}%
\label{fig3}%
\end{figure}

In order to investigate the origin of nonreciprocal PBGs, we present the
reflectivities and transmissivities for Part I (II) considered as an
individual lattice with a length of $0.585$ mm in Figs.~\ref{fig2}(b)
[\ref{fig2}(c)]. The Bragg condition for the Part I lattice is given by
\begin{equation}
\frac{\omega_{p}}{c}(a_{f}\lambda_{0}\overline{n_{f}(x)}+a_{v}\lambda_{0}%
n_{v})=k\pi\text{,} \label{Eq10}%
\end{equation}
with the vacuum speed of light $c$, an arbitrary positive integer $k$, and
average refractive index $\overline{n_{f}(x)}$ for a filled cell. Since
$\overline{n_{f}(x)}>n_{v}=1$, it follows that $(a_{f}+a_{v})\lambda
_{0}\overline{n_{f}(x)}>a_{f}\lambda_{0}\overline{n_{f}(x)}+a_{v}\lambda
_{0}n_{v}$. This implies that the optical path will decrease when some filled
cells are replaced by vacant cells, and then the frequency $\omega_{p}$ should
increase, that is, $\Delta_{p}$ decrease, to satisfy the Bragg condition
again. Therefore, compared to the reflection $R_{\text{II}}$ of the lattice of
Part II [see Fig.~\ref{fig2}(c)], the reflection band $R_{\text{I}}$ of the
lattice of Part I is blue-shifted [see Fig.~\ref{fig2}(b)]. We note the
reflection bands of $R_{r}$ in Fig.~\ref{fig2}(a) and $R_{\text{II}}$ in
Fig.~\ref{fig2}(c) coincide because the right-incident light propagating along
the defective lattice firstly encounters Part II and is then perfectly
reflected. Additionally, $R_{l}$ in Fig.~\ref{fig2}(a) and $R_{\text{I}}$ in
Fig.~\ref{fig2}(b) do except the narrow peak of $R_{l}$ in the shaded regions.
The underlying physics of forming the narrow band is that Part I is
transparent for the left-incident light carrying frequencies in the shaded
region [see $T_{\text{I}}$ in Fig.~\ref{fig2}(b)], and therefore, the light
could pass Part I and be reflected backwards to the left side by Part II [see
$R_{\text{II}}$ in Fig.~\ref{fig2}(c)].

We focused on reflection behaviors at the valley point $\Delta_{p}=4.7$ MHz.
Fig.~\ref{fig3}(a) illustrates that the real and imaginary parts of the
reflection coefficient $r_{_{r}}$ for the right-incident light monotonically
increase as the lattice length increases, which can be used to check the
reflection on the end of the medium. After $L\approx0.25$ mm, they exhibit an
identical tendency towards their respective steady values. However, a
noticeable non-monotonic variation is observed in the real and imaginary parts
of $r_{_{l}}$ for the left-incident light. Due to the more complex periodic
structure of Part I, they tend towards stability only after the light has
entered Part II. As a result, a more extended lattice (approximately $0.75$
mm) is required to attain their steady values. Such a nonreciprocity of
two-side reflections is caused by the lack of conjugacy between complex
transfer matrix elements $M_{d}(1,2)$ and $M_{d}(2,1)$. The underlying physics
is that the insertion of vacant cells on one side breaks the spatial symmetry
of the lattice structure. The differences in the value, width, and position
between the two reflection-band spectra of $R_{l}$ and $R_{r}$ with varying
lengths are intuitively shown in Figs.~\ref{fig3}(b) and (c). Note, as shown
in Fig.~\ref{fig3}(b), the narrow peak of $R_{l}$ becomes visible and
progressively stronger only after $L={\large 0.6}$ mm, indicating the
extension of the lattice into Part II. This also demonstrates that the narrow
peak of $R_{l}$ arises from the cooperation between Part I and Part II.

\begin{figure}[t]
\includegraphics[width=0.5\textwidth]{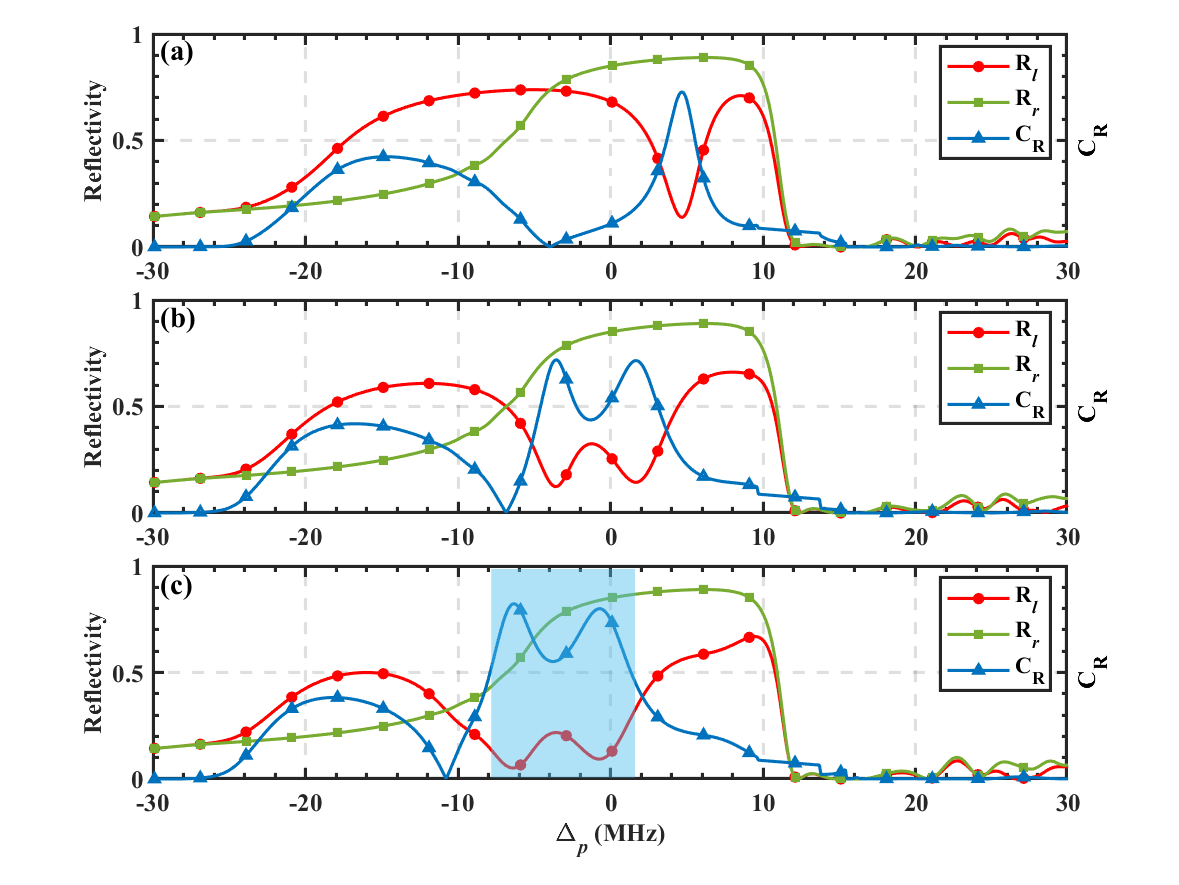}\caption{(Color online)
Reflectivities $R_{l,r}$ and the reflection contrast $C_{R}$ versus probe
detuning $\Delta_{p}$ with (a) $a_{v}/a_{f}=2$, (b) $a_{v}/a_{f}=4$, and (c)
$a_{v}/a_{f}=6$. Here $a_{f}=50$, $p_{1}=10$ and $p_{2}=1500$, other
parameters are the same as in Fig. 1.}%
\label{fig4}%
\end{figure}

As mentioned above, the deviation of the two-side reflection bands is evident
at the valley point, which can be quantified by the reflection contrast, given
by
\begin{equation}
C_{R}=\left\vert \frac{R_{l}-R_{r}}{R_{l}+R_{r}}\right\vert \text{.}
\label{Eq11}%
\end{equation}
According to the Bragg condition for the whole defective lattice
\begin{equation}
\frac{\omega_{p}}{c}\left[  p_{1} (a_{f}\lambda_{0}\overline{n_{f}(x)}%
+a_{v}\lambda_{0} n_{v})+p_{2}\lambda_{0}\overline{n_{f}(x)}\right]
=k\pi\text{,} \label{Eq12}%
\end{equation}
To achieve high-quality nonreciprocal reflection or even unidirectional
reflection, we can design and optimize the structure of the defective lattice
by adjusting $a_{f,v}$ and $p_{1,2}$. Since vacant cells play a critical role,
we change parameters relative to them, for example.

\begin{figure}[t]
\includegraphics[width=0.5\textwidth]{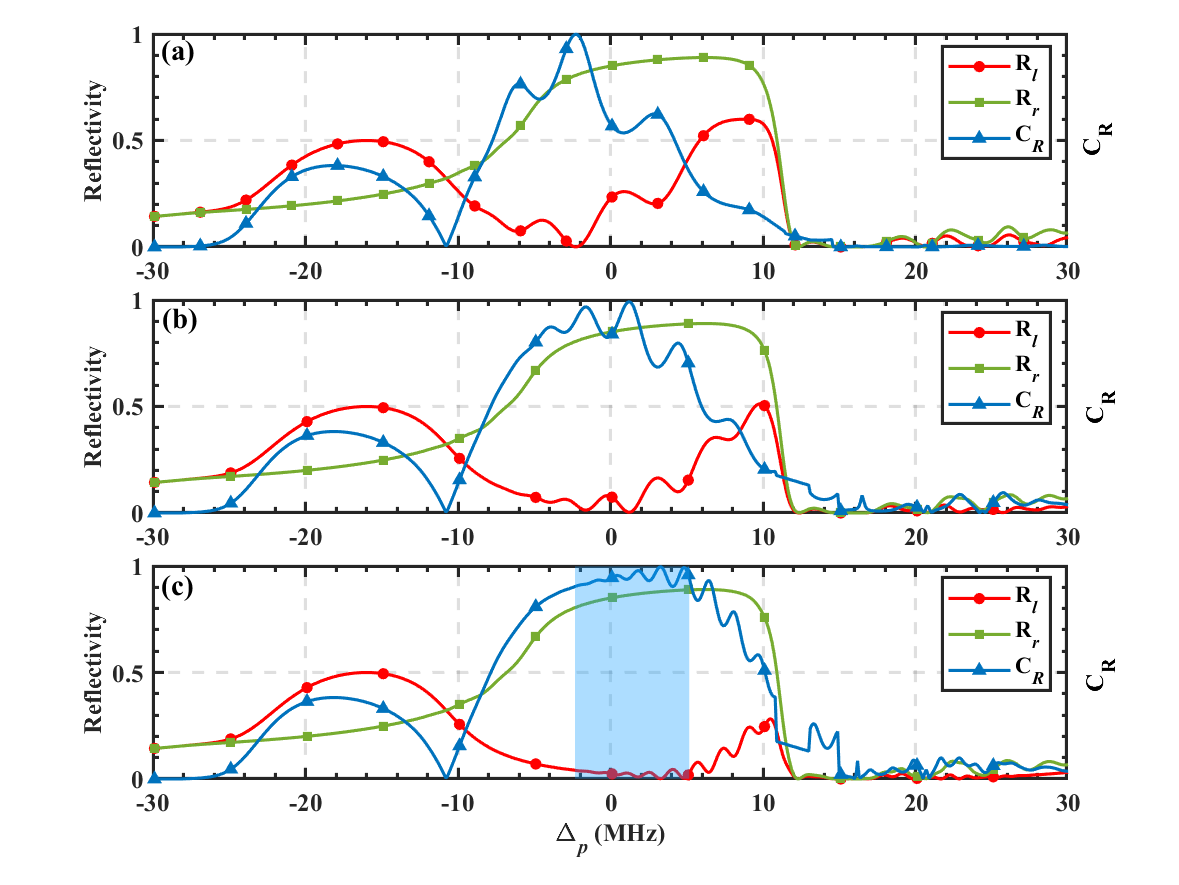}\caption{(Color online)
reflectivities $R_{l,r}$ and the reflection contrast $C_{R}$ versus probe
detuning $\Delta_{p}$ with (a) $p_{1}=15,$ (b) $p_{1}=30$, and (c) $p_{1}=70$.
Here $a_{f}=50$, $a_{v}=100$, and $p_{2}=1500$, other parameters are the same
as in Fig. 1.}%
\label{fig5}%
\end{figure}

Increasing the ratio $a_{v}/a_{f}$ in each period in Part I, accompanied by a
decrease in the refractive index, can lead to a further blue shift for the
wide platform in the LS reflection band. As shown in Fig.~\ref{fig4}, the
valley of the LS reflection band gradually blue-shifts with an increase of
$a_{v}/a_{f}$. Consequently, the narrow peak in the LS reflection band, mainly
due to Part II, widens but does not weaken significantly with the changes in
Part I. However, the wide platform blue-shifts and weakens, gradually leaving
the EIT regime. As $a_{v}/a_{f}$ increases from $2$ to $6$, the band of high
reflection contrast $C_{R}$ gradually blue-shifts, widens to a certain extent
and ultimately produces a broad band of $C_{R}$ over ${\large 50\%}$ as shown
in the shade region in Fig.~\ref{fig4}(c). This allows for broadband
nonreciprocal devices. Especially, with $a_{v}/a_{f}=6$, $C_{R}$ can reach the
maximum $83.50\%$ at $\Delta_{p}=-6.30$ MHz. This indicates the realization of
a perfect unidirectional photonic reflector based on PBGs. Meanwhile, the RS
reflection band remains unchanged. This can be explained further by
considering instances in which the right-incident field propagates through and
is Bragg-scattered by Part II of $1500$ continuous filled cells, which is
sufficient to support a high PBG without any influence from part I; however,
the reflection spectra of LS probe field is primarily altered by Part I and
partially affected by Part II.

To further optimize the nonreciprocal reflection band of high $C_{R}$, one
feasible approach is to suppress the LS reflection band overlapping the RS
one. This can be achieved without compromising the nonreciprocity by adjusting
the number $p_{1}$ of periods of part I. Figure~\ref{fig5} demonstrates that
increasing $p_{1}$ can significantly enhance the perfect nonreciprocal
reflection band, as indicated by high $C_{R}$. With $p_{1}=70$, the
nonreciprocal reflection band of $C_{R}\geq90.0\%$ covers $\Delta_{p}\in$
$[-2$ MHz, $5$ MHz], as shown in the shaded region in Fig.~\ref{fig5}(c). It
is noteworthy that this nonreciprocal band is also within the RS reflection
band of high $R_{r}$ while the LS reflection is effectively restrained.
$C_{R}\simeq100\%$ can be achieved at certain points where $R_{l}\simeq0$.
Hence, such a unidirectional reflector scheme offers some advantages, such as
broadband performance, and high reflectivity of unidirection.

\begin{figure}[t]
\includegraphics[width=0.5\textwidth]{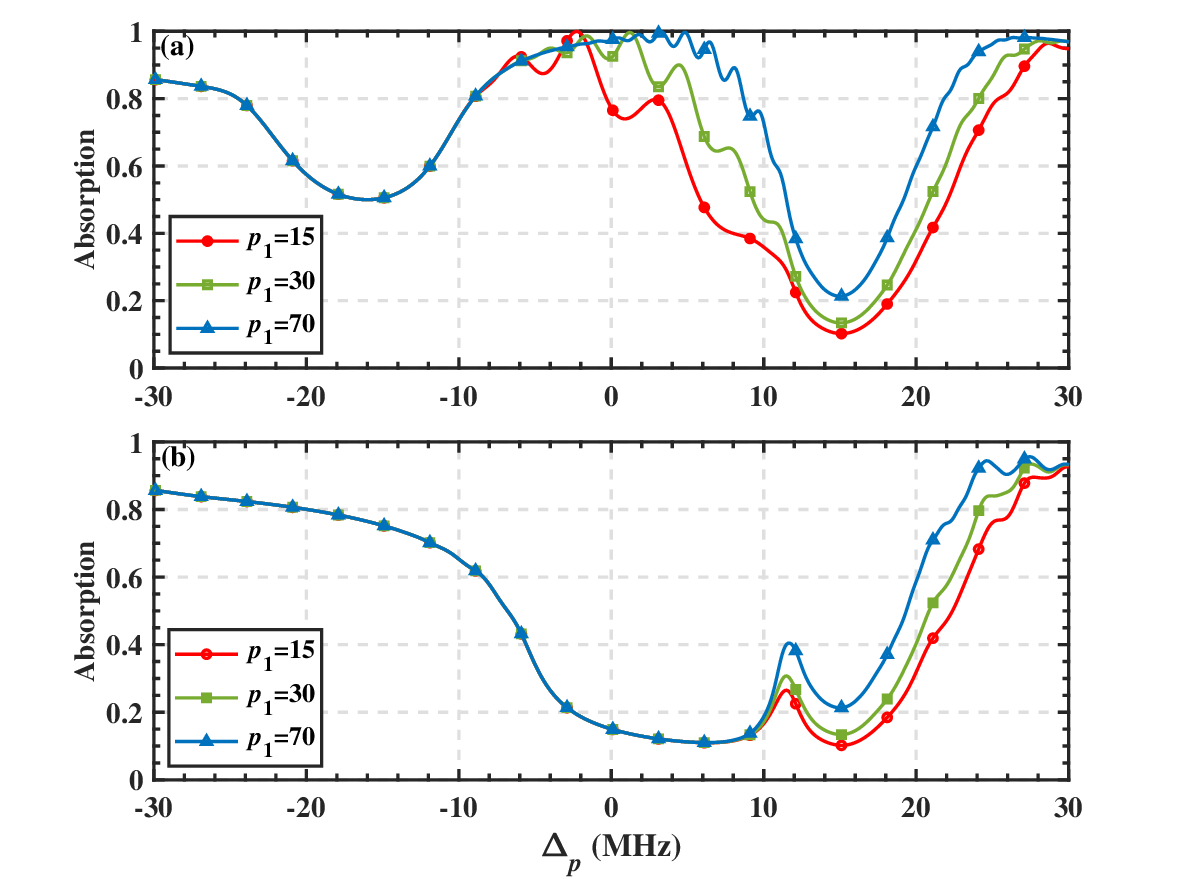}\caption{(Color online) (a)
Absorptivity $A_{l}$ and (b) absorptivity $A_{r}$ versus probe detuning
$\Delta_{p}$ with $p_{1}=15$, $p_{1}=30$, and $p_{1}=70$. Here $a_{f}=50$,
$a_{v}=100$, and $p_{2}=1500$, other parameters are the same as in Fig. 1}%
\label{fig6}%
\end{figure}

Finally, to explain the strong suppression of LS reflection in the perfect
nonreciprocal band, we plot the absorptivities $A_{l}$ and $A_{r}$ for left-
and right-incident fields, respectively, in Figs.~\ref{fig6}. As shown in
\ref{fig6}(a), the absorption for the left-incident field in the perfect
nonreciprocal band gradually increases with the period number $p_{1}$ while
the absorption for the right-incident field remains unchanged. This suggests
that the increasing period number $p_{1}$ can induce one-way absorption
enhancement that results in the complete one-way suppression of reflection.
The main reason for this is that as $p_{1}$ increase, more atoms can be
introduced by the added filled cells, and the left-incident field of frequency
in the narrow peak of $R_{l}$ will experience much more absorption propagating
forward and backward through Part I.

In summary, we propose a unidirectional photonic reflector scheme realized
with a 1D defective atomic lattice of $^{87}$Rb atoms by intentionally
designing some lattice cells to prevent atom trapping. The defective lattice
consists of two parts of periodic structures. Part I, is a periodic structure
where vacant cells appear periodically, while Part II, is a periodic structure
where all cells are filled. The spatial symmetry is broken due to the distinct
positions of the two parts. When considering two parts with the same cells,
the vacant cells induce a blue shift of the reflection band of Part I relative
to the reflection band of Part II. Consequently, the reflection nonreciprocity
arises in such a defective lattice. Unlike the reflection band $R_{r}$ mainly
arising from Part II and a wide platform in the band $R_{l}$ mainly attributed
to Part I, a narrow peak appears in band $R_{l}$ due to the cooperation of
Part I and II. The suppression of the narrow peak of $R_{l}$, by adjusting the
ratio of numbers of vacant and filled cells in each period of Part I and the
period number of Part I, allows us to obtain a broad nonreciprocal reflection
band of high reflection contrast. Our feasible scheme arising from the
breaking of spatial symmetry can be used to study some nonreciprocity
problems. It also gives rise to potential avenues for realizing some
all-optical nonreciprocal circuits and devices, such as optical diodes and
isolators, one-way filters, and routers.

This work is supported by the Hainan Provincial Natural Science Foundation of
China (Grant No. 121RC539) and the National Natural Science Foundation of
China (Grant Nos. 12204137, 12126314, 12126351). This project is also
supported by the specific research fund of The Innovation Platform for
Academicians of Hainan Province Grant (No.YSPTZX202215).

\end{document}